\documentclass[prl,twocolumn, superscriptaddress,amsmath,floats]{revtex4}
\usepackage{txfonts}
\usepackage{amssymb}
\usepackage{graphicx}

\begin{document}

\title{Phosphor induced significant hole-doping in ferropnictide superconductor
BaFe$_2$(As$_{1-x}$P$_x$)$_2$}

\author{Z. R. Ye}\author{Y. Zhang}\author{M. Xu}\author{Q. Q. Ge}\author{F. Chen}\author{Juan Jiang}\author{B. P. Xie}
\affiliation{State Key Laboratory of Surface Physics, Department of Physics,  and Advanced Materials Laboratory, Fudan
University, Shanghai 200433, People's Republic of China}

\author{J. P. Hu}
\affiliation{Department of Physics, Purdue University, West Lafayette, Indiana
47907, USA}

\author{D. L. Feng}\email{dlfeng@fudan.edu.cn}
\affiliation{State Key Laboratory of Surface Physics, Department of Physics, and Advanced Materials Laboratory, Fudan
University, Shanghai 200433, People's Republic of China}

\maketitle

\textbf{The superconductivity in high temperature superconductors
ordinarily arises when doped with hetero-valent ions that introduce
charge carriers \cite{ZXShenRev, La, BaK, BaCo}. However, in ferropnictides, ``iso-valent'' doping, which is
generally believed not to introduce charge carriers, can induce
superconductivity as well \cite{isovalent1, isovalent2, isovalent3,
isovalent5, isovalent6, isovalent7, isovalent9}. Moreover, unlike
other ferropnictides \cite{nodeless1, nodeless2},  the
superconducting gap in BaFe$_2$(As$_{1-x}$P$_x$)$_2$ has been
found to contain nodal lines
\cite{linenode1, linenode2, linenode3}. The exact nature of the
``iso-valent'' doping and nodal gap here are key open issues in
building a comprehensive picture of the iron-based high temperature
superconductors \cite{dxy3, dxy4, dxy5, dxz}. With angle-resolved
photoemission spectroscopy (ARPES), we found that the phosphor
substitution in BaFe$_2$(As$_{1-x}$P$_x$)$_2$ induces sizable amount
of holes into the hole Fermi surfaces, while the
$d_{xy}$-originated band is relatively intact. This overturns the previous
common belief of ``iso-valent'' doping, explains why the phase diagram of
BaFe$_2$(As$_{1-x}$P$_x$)$_2$ is similar to those of the hole-doped
compounds, and rules out theories that explain the nodal gap based on vanishing $d_{xy}$ hole
pocket.}

BaFe$_2$(As$_{1-x}$P$_x$)$_2$ is a rather unique ferropnictide as
its superconductivity is introduced by the iso-valent doping of P
for As  \cite{isovalent5, isovalent2}. Unlike the hetero-valent
doping that alters the carrier concentration in
Ba$_{1-x}$K$_x$Fe$_2$As$_2$, BaFe$_{2-x}$Co$_{x}$As$_2$, or
LaO$_{1-x}$F$_x$FeAs \cite{La, BaK, BaCo}, the iso-valent doping is
often considered not to alter the occupation of the Fe $3d$ bands,
as illustrated by the density functional theory calculations of
BaFe$_2$As$_2$ and BaFe$_2$P$_2$ as well  \cite{isovalent2,
isovalent3}. Yet, surprisingly, it has a similar phase diagram just
like the hetero-valent doped cases: with P doping, spin density wave
(SDW) is suppressed and superconductivity (SC) emerges
\cite{isovalent2}.

Since P anion is smaller than As anion, and thus introduces internal
strain or distortion, i.e. chemical pressure, the superconductivity
introduced by iso-valent doping is associated with the unprecedented
pressure dependence of the superconducting transition temperature
($T_c$) generally observed in iron-based superconductors
\cite{pres1, pres2, pres4, PLa}. In fact, it is the largest among
all superconductors in both relative and absolute scales. For
example, a $T_c$ dependency of 2-4K/GPa and sometimes even 10K/GPa
is observed in BaFe$_2$(As$_{1-x}$P$_x$)$_2$, LaO$_{1-x}$F$_x$FeAs,
etc. \cite{PLa, pres4}; and an increase of $T_c$ from 0 to above
30~K is observed in BaFe$_2$As$_2$ and FeSe under pressure
\cite{pres1, pres2}. However, these remarkable pressure effects are
still far from understood. Theoretically, P doping is predicted to
alter the band structure and Fermi surface topology dramatically,
considering it changes the electron hopping terms \cite{Vildosola,
dxy5}. Particularly, it is predicted that the $d_{z^2}$-based band
would go above the Fermi energy ($E_F$), while the $d_{xy}$-based
band would move down below $E_F$ with P doping. Several theories
further claim that nodes will appear in the superconducting gap when
the $d_{xy}$ hole Fermi pocket disappears \cite{dxy3, dxy4, dxy5}.

\begin{figure}[b]
\includegraphics[width=8.7cm]{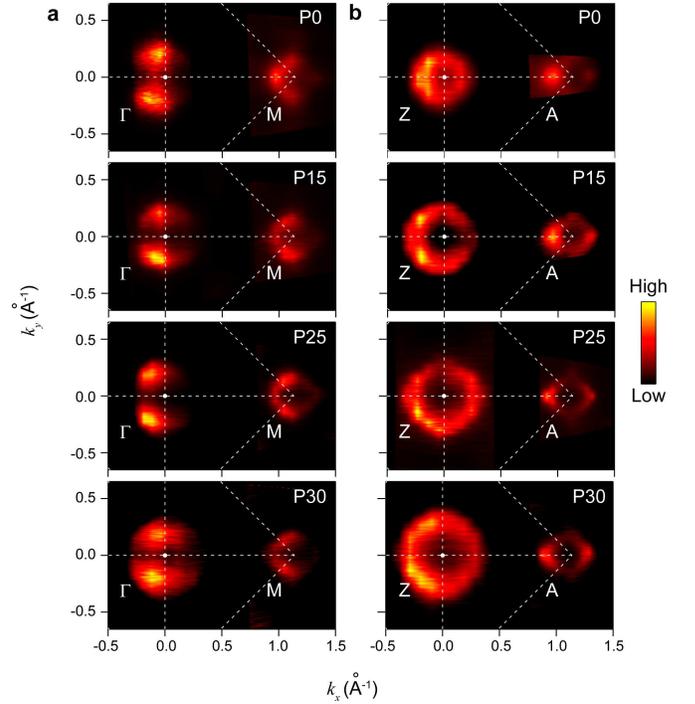}
\caption{\textbf{The doping dependence of the photoemission
intensity map in BaFe$_2$(As$_{1-x}$P$_x$)$_2$}. \textbf{a},
Photoemission intensity maps around $\Gamma$ and M ($k_z$~=~0) in
BaFe$_2$(As$_{1-x}$P$_x$)$_2$ for $x$~=~0, 0.15, 0.25, 0.30
respectively, named as P0, P15, P25, and P30 hereafter.  \textbf{b},
is the same as \textbf{a}, but taken around Z and A
($k_z$~=~2$\pi$/c). Data around $\Gamma$, M, Z, and A were taken at
22~eV, 26~eV, 33~eV, and 17~eV respectively. All data were measured
in the paramagnetic state at 150~K, 110~K, 75~K, and 40~K for P0,
P15, P25 and P30 respectively to avoid the complications from the
electronic structure reconstruction in the SDW and superconducting
states \cite{YangBF, ZhangBK}.} \label{mapping}
\end{figure}

\begin{figure*}[t]
\includegraphics[width=18cm]{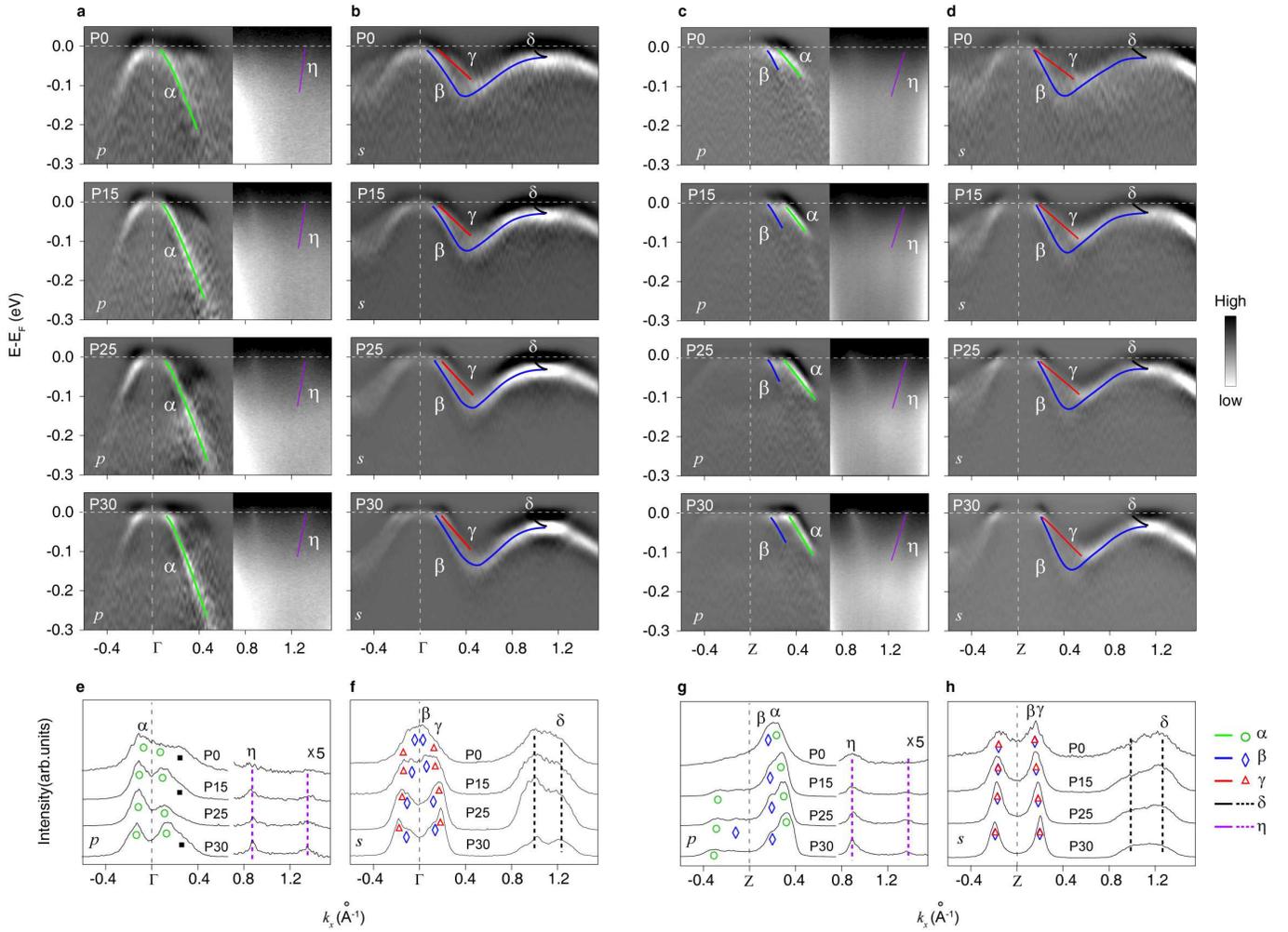}
\caption{\textbf{The doping and polarization dependence of the
photoemission data}. \textbf{a}, Left side: the second derivative
with respect to energy of the photoemission intensity taken along
$\Gamma$-M direction with 118~eV photons for P0 and P15, and 120~eV
photons for P25 and P30 in the $p$ polarization; Right side:
photoemission intensity highlighting the fast dispersing weak
features. \textbf{b}, is the same as the left side of \textbf{a},
but taken in the $s$ polarization. \textbf{e},\textbf{f}, the
corresponding momentum distribution curves (MDC's) near $E_F$ of
\textbf{a},\textbf{b}. Note that the right part of \textbf{e} is in
an expanded intensity scale. The peaks marked by black solid squares
in \textbf{e} are contributed by the residual spectral weight of the
band below $E_F$ with the $d_{z^2}$ orbital (see supplementary
information for details). \textbf{c}, \textbf{d}, \textbf{g}, and
\textbf{h}, are the same as \textbf{a}, \textbf{b}, \textbf{e}, and
\textbf{f} respectively, but taken along $Z$-A direction with 100~eV
photons. Solid lines are band dispersion, as determined by MDC peaks
or the minima of the second derivative of photoemission intensity
with respect to energy. Only half of data are overlayed with
determined dispersions for a better view of the data.}
\label{orbital}
\end{figure*}

Figure~\ref{mapping} examines the dependence of the Fermi surfaces
on the P concentration in a series of BaFe$_2$(As$_{1-x}$P$_x$)$_2$,
where the  photoemission intensity maps near $E_F$ are shown for two
$k_z$'s. The features  at the zone center ($\Gamma$ and Z) are hole
pockets, and those at the zone corner (M and A) are electron pockets
\cite{YangBF, Zhangorbital}. As P doping increases, the size of the
hole pockets increase significantly, while the electron pockets show
negligible doping dependence. This indicates that the P doping could
induce extra holes into the system, contradicting to the ordinary
picture of iso-valent doping.

\begin{figure*}[t]
\includegraphics[width=15cm]{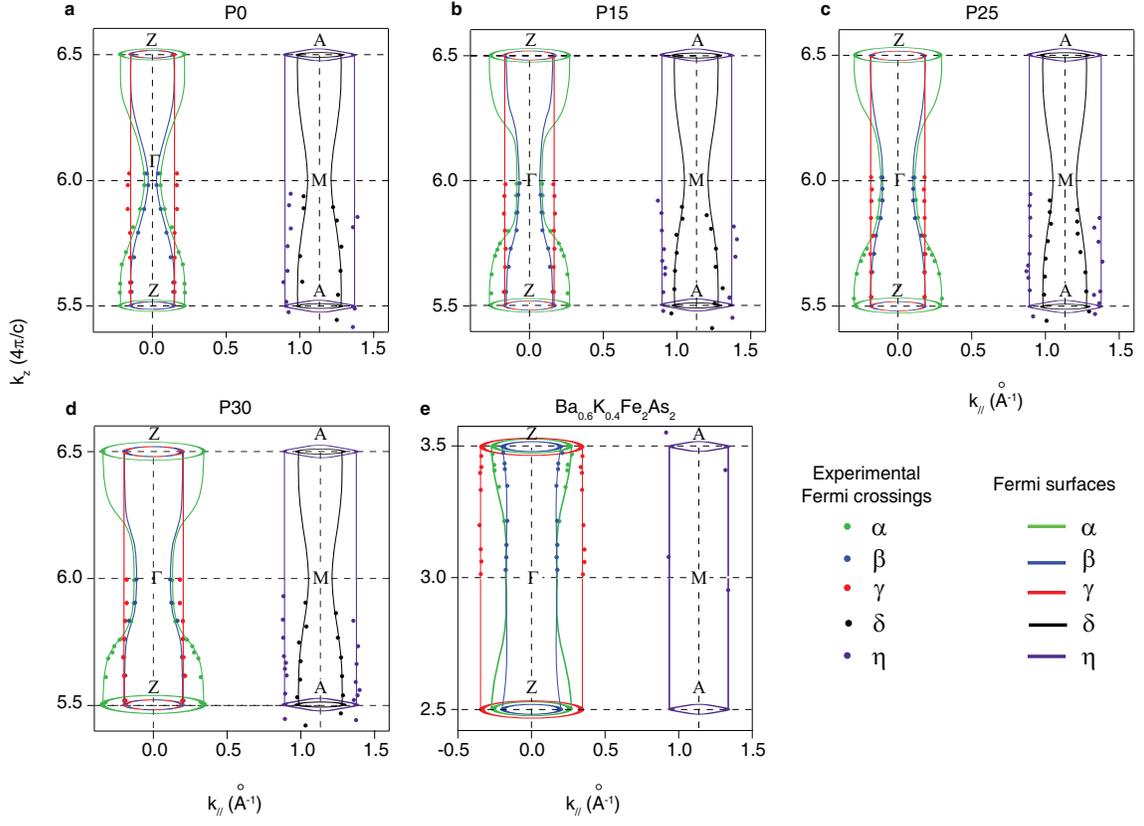}
\caption{\textbf{The experimental  Fermi surface cross-sections in
Z-$\Gamma$-M-A plane}. \textbf{a}-\textbf{d},  Fermi surfaces of P0,
P15, P25, and P30  respectively determined based on the experimental
Fermi crossings (dots). \textbf{e}, Fermi surface of
Ba$_{0.6}$K$_{0.4}$Fe$_2$As$_2$ \cite{ZhangBK}.} \label{phd}
\end{figure*}

To understand such extraordinary P doping effect, more detailed band
structures near $E_F$ are examined with both $s$ and $p$
polarizations of the incoming photons. Since the low-lying
electronic structures are mostly made of Fe-$3d$ orbitals with
specific even or odd spatial geometry, the photoemission intensity
of the even (or odd) component of a band is only detectable with the
$p$ or $s$ polarized light
 \cite{Zhangorbital} (see supplementary information for details).
Fig.~\ref{orbital}a shows the photoemission data taken around
$\Gamma$ with the $p$ polarization. One could resolve a fast
dispersing $\alpha$ band. According to previous studies, the
$\alpha$ band corresponds to the even $d_{xz}$ orbital
 \cite{Zhangorbital}. In the $s$ polarization (Fig.~\ref{orbital}b), two
bands,  $\beta$ and $\gamma$, could be observed. The $\beta$ band
could be assigned to the odd $d_{yz}$ orbital. Although $\beta$ is
almost degenerate with  $\alpha$  near $E_F$, it bends back towards
$E_F$ at about 120~meV, instead of further dispersing to higher
binding energies. The $\gamma$ band is made of the $d_{xy}$ orbital,
which shows much weaker intensity than $\beta$  due to its small
matrix element near the $\Gamma$ point  \cite{Zhangorbital}. Around
the Z point, however, one could resolve two bands in the $p$
polarization (Fig.~\ref{orbital}c). The inner band is contributed by
the even component of the $\beta$ band, since it shows the identical
Fermi crossing and band dispersion with the odd component of the
$\beta$ band observed in the $s$ polarization (Fig.~\ref{orbital}d).
The outer one is the $\alpha$ band, whose Fermi momentum is enlarged
significantly from $\Gamma$ to $Z$, indicating its remarkably strong
$k_z$ dispersion. On the other hand, the Fermi momentum of $\beta$
moves outward only slightly and becomes degenerate with the $\gamma$
band around Z.

Two electron-like bands, $\delta$ and $\eta$, are clearly
distinguished in the $s$ and $p$ polarization separately around the
zone corner, consistent with previous studies \cite{Zhangorbital}.
Because of the weak intensity, the $\eta$ band is presented with the
photoemission intensity data in Fig.~\ref{orbital}a,c, and the Fermi
crossings of these two bands together with others could be traced
from the corresponding momentum distribution curves (MDC's) near
$E_F$ in Fig.~\ref{orbital}e-h. It is clear that the Fermi crossings
of $\delta$ and $\eta$ show negligible doping dependence, while the
Fermi momenta of $\alpha$, $\beta$ and $\gamma$ increase with P
doping. Particularly, the $\alpha$  Fermi crossings  expand much
more exceptionally than the others around Z.

\begin{figure*}[t]
\includegraphics[width=15cm]{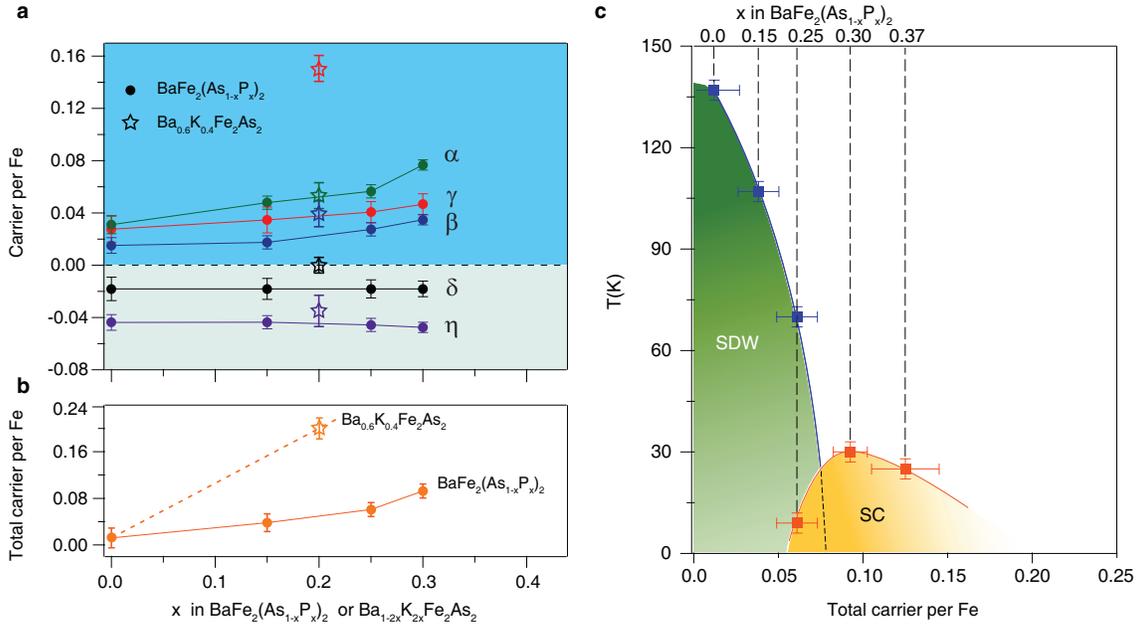}
\caption{\textbf{Summary of the band-specific carrier density, and
phase diagram of BaFe$_{2}$(As$_{1-x}$P$_x$)$_2$}. \textbf{a}, The
carrier density contributed by individual bands, and \textbf{b}, the
total carrier density  in BaFe$_{2}$(As$_{1-x}$P$_x$)$_2$ and
Ba$_{0.6}$K$_{0.4}$Fe$_2$As$_2$.  \textbf{c}, The phase diagram of
BaFe$_{2}$(As$_{1-x}$P$_x$)$_2$ with respect to the total carrier
density. The error bars of carrier density per Fe in \textbf{a},
\textbf{b}, and \textbf{c} are due to the uncertainty in estimating
the 3D Fermi surface size. The error bars of the transition
temperatures in \textbf{c} are obtained from the transport
measurement.}\label{sum}
\end{figure*}

The different behaviors of the band structure between $\Gamma$ and Z
suggest a strong three-dimensional (3D) character of the electronic
structure. Through photon energy dependence of the MDC's in both the
$s$ and $p$ polarizations (see supplementary information for
details), we could determine the Fermi crossings of individual
bands. The resulting Fermi surface cross-sections in the $k_x$-$k_z$
plane are shown in Fig.~\ref{phd}a-d. In all cases, the two electron
pockets $\delta$ and $\eta$ at the zone corner show a weak 3D
character and negligible doping dependence \cite{isovalent3}. The
$\alpha$ and $\beta$ hole pockets increase from $\Gamma$ to Z, while
the $\gamma$ pockets show almost no $k_z$ dependence, which is
consistent with the two-dimensional character of the $d_{xy}$
orbital \cite{Zhangorbital}. With P doping, $\beta$ and $\gamma$
Fermi surfaces expand slightly, keeping their shapes basically
unchanged. However, the curvature of the $\alpha$ Fermi surface
changes significantly, which suggests a strong enhancement of the 3D
character of the $\alpha$ band due to P doping. While previous
quantum oscillation \cite{isovalent3} and ARPES measurements
\cite{shin} were focused on limited regions of the Brillouin zone,
our data over the entire Brillouin zone give direct evidence for the
phosphor induced hole doping. Moreover, the three hole pockets
observed here complement the previous ARPES study \cite{isovalent6},
where only two hole pockets were observed due to its particular
experimental setup. Furthermore, our results provide compelling
evidence that the $d_{xy}$ band still forms the $\gamma$ Fermi
surface near $E_F$, which is inconsistent with the prediction of
band calculations \cite{dxy3, dxy4, dxy5}. Consequently, our data
rule out all the theories that explain nodal gap in
BaFe$_2$(As$_{1-x}$P$_x$)$_2$ based on the vanishing $d_{xy}$ Fermi
surface \cite{dxy3, dxy4, dxy5}.

The hole doped by phosphor in BaFe$_2$(As$_{1-x}$P$_x$)$_2$ is
unexpected. As a comparison, Fig.~\ref{phd}e shows the Fermi surface
cross-section of the optimally hole-doped
Ba$_{0.6}$K$_{0.4}$Fe$_2$As$_2$ with a $T_c$ as high as 38~K
\cite{ZhangBK}, where the doped potassium is outside the FeAs layer
and should not induce much chemical pressure in the FeAs layer.
Compared with the undoped case (Fig.~\ref{phd}a), the hole doping in
Ba$_{0.6}$K$_{0.4}$Fe$_2$As$_2$  enlarges the hole pockets, and
shrinks the electron pockets as expected in a rigid band picture,
where the $\delta$ electron pocket even diminishes. On the other
hand, the different behaviors of electron and hole pockets in
BaFe$_2$(As$_{1-x}$P$_x$)$_2$ manifest its strong non-rigid band
nature, which is most likely caused by chemical pressure. More
specifically, the $\gamma$ pocket is much smaller, and the electron
pockets ($\eta$ and $\delta$) are lager in P30 than that in
Ba$_{0.6}$K$_{0.4}$Fe$_2$As$_2$. Another remarkable difference
between P30 and Ba$_{0.6}$K$_{0.4}$Fe$_2$As$_2$ is the $\alpha$
Fermi surface, which exhibits a strong $k_z$ dependence, and becomes
the outmost hole pocket around Z in P30. Recent theoretical study
has proposed that, with the increasing of the phosphor doping, the
weight of $d_{z^2}$ orbital could shift up towards $E_F$, and
strongly mixed into the bands around Z, which could result in the
nodal superconducting gap \cite{dxz}. Since the mixture of $d_{z^2}$
might enhance $k_z$ dependence, our results seem to favor this
scenario. Thus, it would be elucidative to study the $d_{z^2}$
component in $\alpha$, and look for nodes in the superconducting gap
on the $\alpha$ Fermi surface around $Z$. We leave this for further
studies.

The charge carrier doping can be estimated from the Fermi surface
volume. Fig.~\ref{sum}a summarizes the carriers per Fe contributed
by each band. The hole carriers in the $\alpha$, $\beta$, and
$\gamma$ bands increase simultaneously with P doping, with the
biggest contribution from the $\alpha$ band. The electron carriers
in the $\delta$ band remain the same, while  those of the $\eta$
band slightly increase with P doping. Fig.~\ref{sum}b plots the
total net carrier per Fe obtained by summing up the contributions
from all these five bands. The doping of holes is nearly
proportional to the P doping, and one phosphor substation could
induce about 0.3 holes per iron. One possible explanation for such a
hole doping comes from the balance between bond length and anion
electron negativity. When the bond length is shorter, an anion would
take less electrons away from the cation of a covalent bond. For
BaFe$_2$As$_2$ and BaFe$_2$P$_2$, the Fe-As bond length is
2.3980~$\AA$, and Fe-P bond length is 2.2614~$\AA$ respectively
\cite{isovalent1}. The electronegativity of P (2.19 in Pauling's
scale) is larger than that of As (2.18) \cite{neg}. The smaller Fe-P
bond length and the larger P electronegativity make the charge
distribution on the Fe-P covalent bond similar to that of Fe-As.
Therefore, holes and electrons in both systems are balanced.
However, for light P-doped case, the P substitution could induce
local lattice distortion and the Fe-P bond length is larger than
that in BaFe$_2$P$_2$ (2.2614~$\AA$). As a result, P would take more
electrons away from Fe, and the localization of electrons at
phosphor sites induces the hole carrier. On the other hand, when the
doping is approaching the BaFe$_2$P$_2$ end, the hole doping would
predictably decrease.

For comparison, the carrier density distribution in
Ba$_{0.6}$K$_{0.4}$Fe$_2$As$_2$ is shown by the pentacles in
Fig.~\ref{sum}a,b, where the $\gamma$ band contributes the most
holes into the system rather than the $\alpha$ band in
BaFe$_2$(As$_{1-x}$P$_x$)$_2$. The total carrier density in
Ba$_{0.6}$K$_{0.4}$Fe$_2$As$_2$ is 0.208$\pm$0.010 holes per Fe, as
expected from the K concentration. That is, the hole doping in
Ba$_{0.6}$K$_{0.4}$Fe$_2$As$_2$ is due to the charge transfer
between the FeAs layer and the potassium layer, instead of the
localization of electrons in BaFe$_2$(As$_{1-x}$P$_x$)$_2$.
Fig.~\ref{sum}c gives the phase diagram with respect to the total
carrier per Fe in BaFe$_2$(As$_{1-x}$P$_x$)$_2$. It greatly
resembles the K-doped case, except that  $T_c$  reaches its maximum
at 0.09 holes per Fe, instead of 0.2 holes per Fe in
Ba$_{1-x}$K$_x$Fe$_2$As$_2$. It suggests that the rise of
superconductivity in BaFe$_2$(As$_{1-x}$P$_x$)$_2$  is not a pure
doping effect. Both the doping effect and the
chemical-pressure-induced non-rigid band behavior described above
should be considered to explain the phase diagram of ``iso-valent''
doping in iron-based superconductors.

Recently, it has been demonstrated that the chemical and physical
pressure are rather equivalent in BaFe$_2$(As$_{1-x}$P$_x$)$_2$, as
the arbitrary combination of both phosphor doping and physical
pressure can reproduce the same phase diagram  \cite{pres4}. Based
on our results, it is not unnatural to speculate that the drastic
physical pressure effect in iron-based superconductors is likely due
to the similar electronic structure changes. However, as the
physical pressure does not introduce carriers, both electron and
hole pockets would vary simultaneously.

To summarize, the iso-valent ionic picture is oversimplified which
could not describe the phosphor doping in iron-based
superconductors. Our results highlight that both chemical pressure
effect on the band structure, and a new route of carrier doping
should play important roles in understanding the various distinct
properties of BaFe$_2$(As$_{1-x}$P$_x$)$_2$. Particularly, the
$d_{xy}$ band does not sink below $E_F$ even at the optimally doped
sample, which disproves the theories that tried to explain the nodal
gap in BaFe$_2$(As$_{1-x}$P$_x$)$_2$ based on the absence of the
$d_{xy}$ Fermi surface. More importantly, the phenomenologies in
BaFe$_2$(As$_{1-x}$P$_x$)$_2$  are united with those by carrier
doping to some extent; and a possible explanation of the remarkable
pressure effects on $T_c$ in iron-based superconductors is given.

\textbf{Methods:} High quality BaFe$_{2}$(As$_{1-x}$P$_x$)$_2$
($x$~=~0, 0.15, 0.25, 0.30) single crystals were synthesized without
flux. Ba, FeAs and FeP were mixed with the nominal compositions,
loaded into an alumina tube, and then sealed into a stainless steel
crucible under the Ar atmosphere. The entire assembly was heated to
1673~K  and kept for 12~h or longer, and then slowly cooled  down to
1173~K at the rate of 4~K/h before shutting off the power. Shiny
platelet crystals as large as 2$\times$ 2$\times$ 0.05~mm$^3$ were
obtained with residual resistivity ratio of about 10(see
supplementary information for details). The P concentrations, $x$,
were confirmed by an energy dispersive X-ray (EDX) analysis. The
ratio of Ba, Fe, and (AsP) is about 1:2.1:1.9 in all samples. The
spin density wave (SDW) transition temperatures were 137~K, 100~K,
and 65~K in x~=~0 (P0), x~=~0.15 (P15), and x~=~0.25 (P25) samples
respectively. Superconductivity were observed in P25 and x~=0.30
(P30) samples with $T_c$~=~ 9~K and 30~K respectively. Mixed
polarization data were taken at the Beamline 5-4 of Stanford
Synchrotron Radiation Lightsource (SSRL). Polarization-dependence
data were taken at the SIS beamline of the Swiss Light Source (SLS).
All the data were taken with Scienta electron analyzers, the overall
energy resolution was 15-20~meV at SLS or 7-10~meV at SSRL depending
on the photon energy, and the angular resolution was 0.3 degree. The
samples were cleaved \textit{in situ}, and measured under
ultra-high-vacuum of 5$\times$10$^{-11}$\textit{torr}.

\textbf{Acknowledgement: }This work is supported in part by the
National Science Foundation of China, Ministry of Education of
China, Science and Technology Committee of Shanghai Municipal, and
National Basic Research Program of China (973 Program)  under the
grant Nos. 2011CB921802 and 2011CBA00102. SSRL is operated by the US
DOE, Office of Basic Energy Science, Divisions of Chemical Sciences
and Material Sciences. We thank Dr. D.H.Lu for his assistance at SSRL, and Dr. X.Y.Cui and Dr. M.Shi for their assistance at SLS.



\end{document}


\title{Supplementary information for:

Phosphor induced significant hole-doping in ferropnictide
superconductor BaFe$_2$(As$_{1-x}$P$_x$)$_2$}

\author{Z. R. Ye}\author{Y. Zhang}\author{M. Xu}\author{Q. Q. Ge}\author{F. Chen}\author{Juan Jiang}\author{B. P. Xie}
\affiliation{State Key Laboratory of Surface Physics, Department of
Physics,  and Advanced Materials Laboratory, Fudan University,
Shanghai 200433, People's Republic of China}

\author{J. P. Hu}
\affiliation{Department of Physics, Purdue University, West
Lafayette, Indiana 47907, USA}

\author{D. L. Feng}\email{dlfeng@fudan.edu.cn}
\affiliation{State Key Laboratory of Surface Physics, Department of
Physics, and Advanced Materials Laboratory, Fudan University,
Shanghai 200433, People's Republic of China}

\maketitle

\section{Sample Description}

High quality BaFe$_{2}$(As$_{1-x}$P$_x$)$_2$ ($x$~=~0, 0.15, 0.25,
0.30) single crystals were synthesized with no flux method. Shining
platelet crystals as large as 2$\times$ 2$\times$ 0.05~mm$^3$ were
obtained as shown in the inset of Fig.~S1a. The resistivity of
BaFe$_2$(As$_{0.7}$P$_{0.3}$)$_2$ single crystal shows a sharp
superconducting transition at 30~K, with the transition width less
than 0.5~K (Fig.~S1a). The room temperature-to-residual resistivity
ratio is $\approx$ 10.4, indicating minimal disorder and impurity of
the single crystal. The magnetic susceptibility as a function of
temperature was measured in a 20~Oe magnetic field (Fig.~S1b). A
sharp drop in the zero field cooling (ZFC) susceptibility,
indicating the magnetic onset of superconductivity, appears at 30K,
which is consistent with the zero-resistivity temperature in
Fig.~S1a. The ZFC and field cooling (FC) curves indicate a bulk
superconductivity nature and high quality of the crystals. The phase
diagram of BaFe$_{2}$(As$_{1-x}$P$_x$)$_2$ is shown in Fig.~S2

\begin{figure*}[h]
\includegraphics[width=15cm]{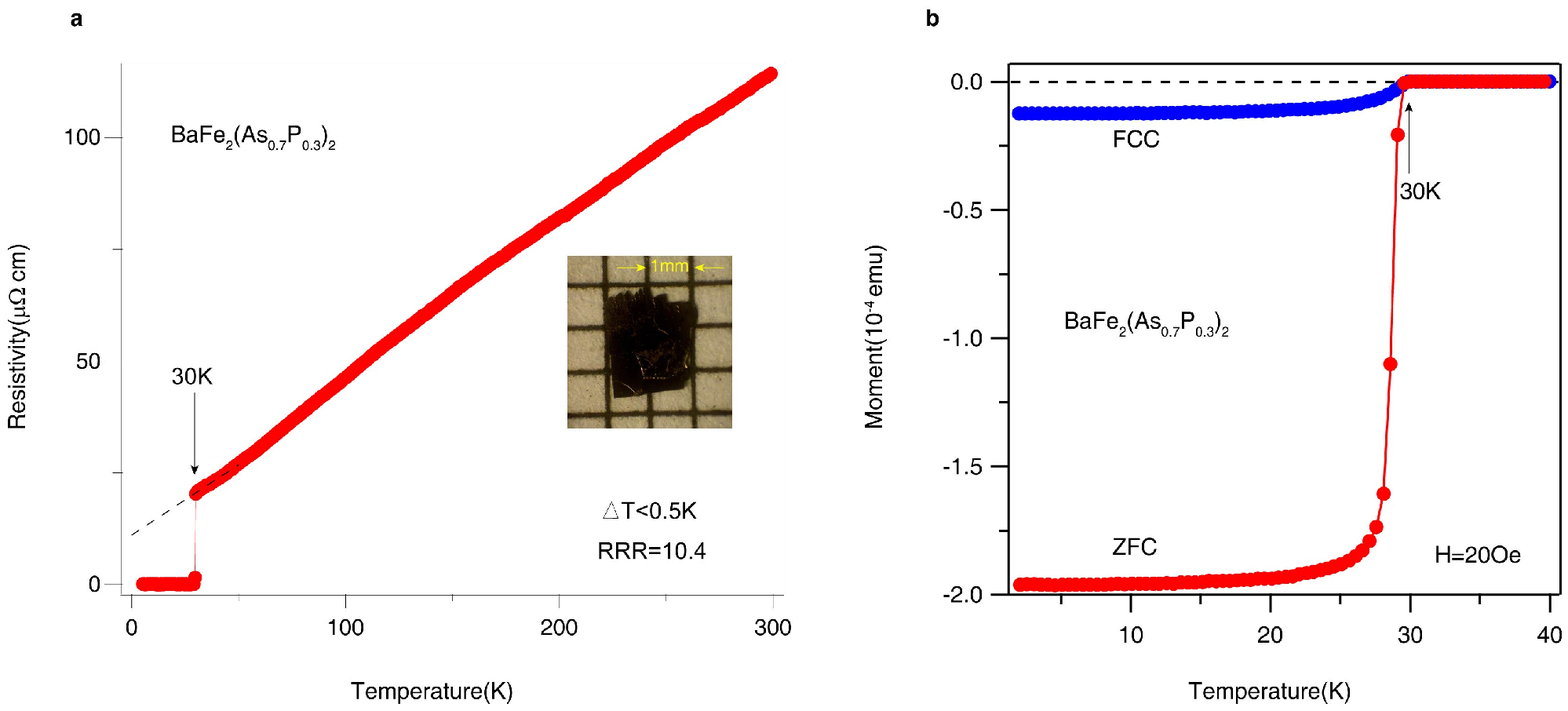}
\caption{\textbf{Electrical resistivity and magnetic susceptibility
 of BaFe$_2$(As$_{0.7}$P$_{0.3}$)$_2$}. \textbf{a}, Temperature dependence of electrical resistivity.
The inset is the photography of the
BaFe$_2$(As$_{0.7}$P$_{0.3}$)$_2$ single crystal. \textbf{b},
Temperature dependence of magnetic susceptibility in a 20~Oe
magnetic field.} \label{trans}
\end{figure*}

\newpage
\begin{figure}[h]
\includegraphics[width=8cm]{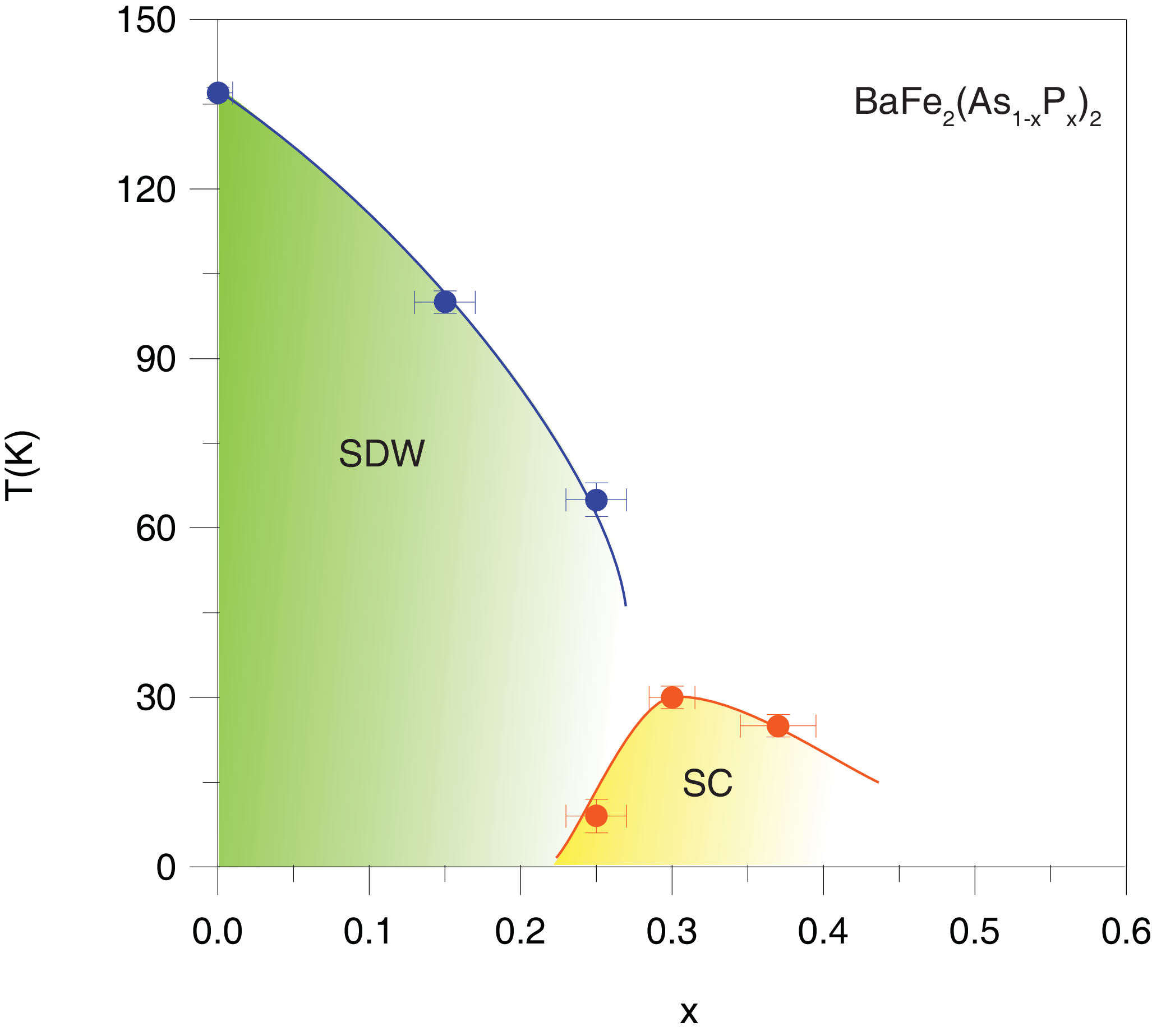}
\caption{\textbf{Phase diagram of BaFe$_2$(As$_{1-x}$P$_x$)$_2$ with
respect to the P doping.} The P concentrations, $x$, were confirmed
by an energy dispersive X-ray (EDX) analysis. With P doping, the
spin density wave (SDW) is suppressed and the superconductivity (SC)
emerges. The SDW transition temperatures are 137~K, 100~K, and 65~K
in x~=~0 (P0), x~=~0.15 (P15), and x~=~0.25 (P25) samples
respectively. Superconductivity is observed in P25, x~=0.30 (P30),
and x~=0.37 samples with $T_c$~=~ 9~K, 30~K, and 24.5~K
respectively.} \label{phase}
\end{figure}

\newpage

\section{Experimental Setup}

The polarization-sensitivity of the orbitals in angle-resolved photoemission spectroscopy (ARPES) is a powerful
tool to identify the orbital characters of band structure
\cite{ZXShenRev, Zhangorbital}. The matrix element of the
photoemission process can be described by
$${|M_{f,i}^{\bf{k}}|^2\propto{\rm{|}}\langle \phi _f^{\bf{k}}
|\bf{\hat{\varepsilon}}\cdot{\bf{r}}|\phi _i^{\bf{k}} \rangle
|^2}$$, where $\bf{\hat{\varepsilon}}$ is the unit vector of the
electric field of the light \cite{ZXShenRev}. For high
kinetic-energy photoelectrons, the final-state wavefunction ${\phi
_f^{\bf{k}}}$ can be approximated by a plane-wave state
${e^{i{\bf{k}\cdot\bf{r}}}}$ with $\bf{k}$ in the mirror plane as
plotted in Fig.~S3c. Consequently, it is always even with respect to
the mirror plane. For the $p$ (or $s$) experimental geometry in
Fig.~S3c, because $\bf{\hat{\varepsilon}}$ is parallel (or
perpendicular) to the mirror plane,
$\bf{\hat{\varepsilon}}\cdot{\bf{r}}$ is even (or odd). Therefore,
to give finite photoemission matrix element,  \textit{i.e.} to be
observable, the initial state $\phi _i^{\bf{k}}$ has to be even (or
odd) in the $p$ (or $s$) geometry.

Considering the spatial symmetry of the $3d$ orbitals (Fig.~S3d),
when the analyzer slit is along the high symmetry direction of the
sample, the photoemission signal of certain orbitals would appear or
disappear by specifying the polarization directions as summarized in
Table~\ref{t1}. For example, with respect to the mirror plane formed
by direction \#1 and sample surface normal (or the $xz$ plane), the
even orbitals ($d_{xz}$, $d_{z^2}$, and $d_{x^2-y^2}$) and the odd
orbitals ($d_{xy}$ and $d_{yz}$) could be only observed in the $p$
and $s$ geometry respectively. Note that, $d_{xz}$ and $d_{yz}$ are
not symmetric with respect to the mirror plane defined by direction
\#2 and surface normal, thus could be observed in both the $p$ and
$s$ geometries.

\begin{table}[b]
\caption{\textbf{The possibility to detect $3d$ orbitals along two
high symmetry directions in the $p$ and $s$ geometry by
polarization-dependent APRES.}}
\begin{tabular}{@{\vrule height 10.5pt depth 4pt  width0pt}ccccccc}
High-symmetry & Experimental &\multicolumn{5}{c}{$3d$ orbitals}\\
\cline{3-7} direction & geometry & $d_{xz}$ & $d_{x^2-y^2}$ &
$d_{z^2 }$&$d_{yz}$&$d_{xy}$\\ \hline
\#1 $\Gamma$(Z)-M(A)  & $p$ &$\surd$&$\surd$&$\surd$& & \\
& $s$ & & & &$\surd$&$\surd$ \\ \hline
\#2 $\Gamma$(Z)-X(R)  & $p$ &$\surd$& &$\surd$ &$\surd$ &$\surd$ \\
& $s$ &$\surd$&$\surd$& &$\surd$ & \\ \hline
\end{tabular}    \label{t1}
\end{table}

\newpage
\begin{figure*}[h]
\includegraphics[width=14cm]{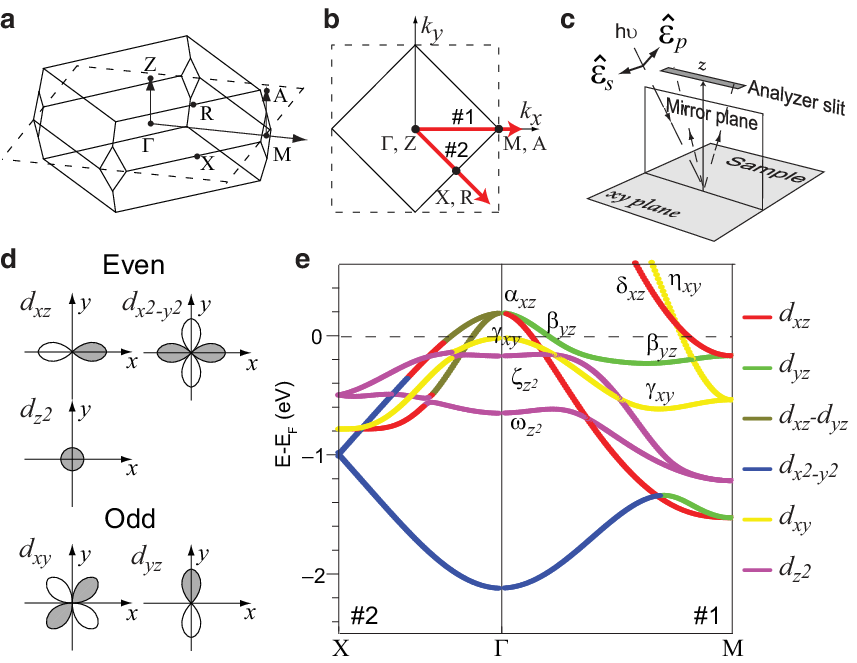}
\caption{\textbf{The experimental setup and definitions.}
\textbf{a}, The illustration of the Brillouin zone of
BaFe$_2$As$_2$. \textbf{b}, Two-dimensional plot of the simplified
Brillouin zone (solid line) and the unfolded Brillouin zone (dashed
line). \textbf{c}, Experimental setup for polarization-dependent
ARPES. For the $p$ (or $s$) experimental geometry, the electric
field direction of the incident photons $\bf{\hat{\varepsilon}_{p}}$
(or $\bf{\hat{\varepsilon}_{s}}$)  is parallel (or perpendicular) to
the mirror plane defined by the analyzer slit and the sample surface
normal. \textbf{d},  Illustration of the spatial symmetry of the
$3d$ orbitals with respect to the mirror plane formed by surface
normal and cut \#1 in panel b, \textit{i.e.} the $xz$ plane.
\textbf{e}, A typical orbital assignment of bands of iron pnictide
as calculated in Ref.~[3].} \label{orbital}
\end{figure*}

\newpage

\section{$d_{z^2}$ orbital}

Besides of the fast dispersing $\alpha$ band, another peak-like
feature could be observed in the momentum distribution curves
(MDC's) in the inset of Fig.~S4a, as pointed out by an arrow. Note
that, the $d_{z^2}$ orbital contributes to one band ($\zeta$), which
is 200~meV below $E_F$ (Fig.~S3e) \cite{Zhangorbital}. This band
disperses towards the $\alpha$ band, resulting in an enhancement of
the spectral weight at the band crossings (Figs.~S4a, b). The
resulted intense spectral weight would extend some residual weight
towards $E_F$ (Fig.~S4c) due to its broad peak width, which forms
the non-dispersing peaks in MDC's as marked by the solid squares in
Fig. ~S4d. Therefore, the peak-like feature ($\zeta$) observed in
the $p$ polarization at $E_F$ is not the Fermi crossing of band, but
the residual spectral weight of $\zeta$ well below $E_F$.

\begin{figure*}[h]
\includegraphics[width=14cm]{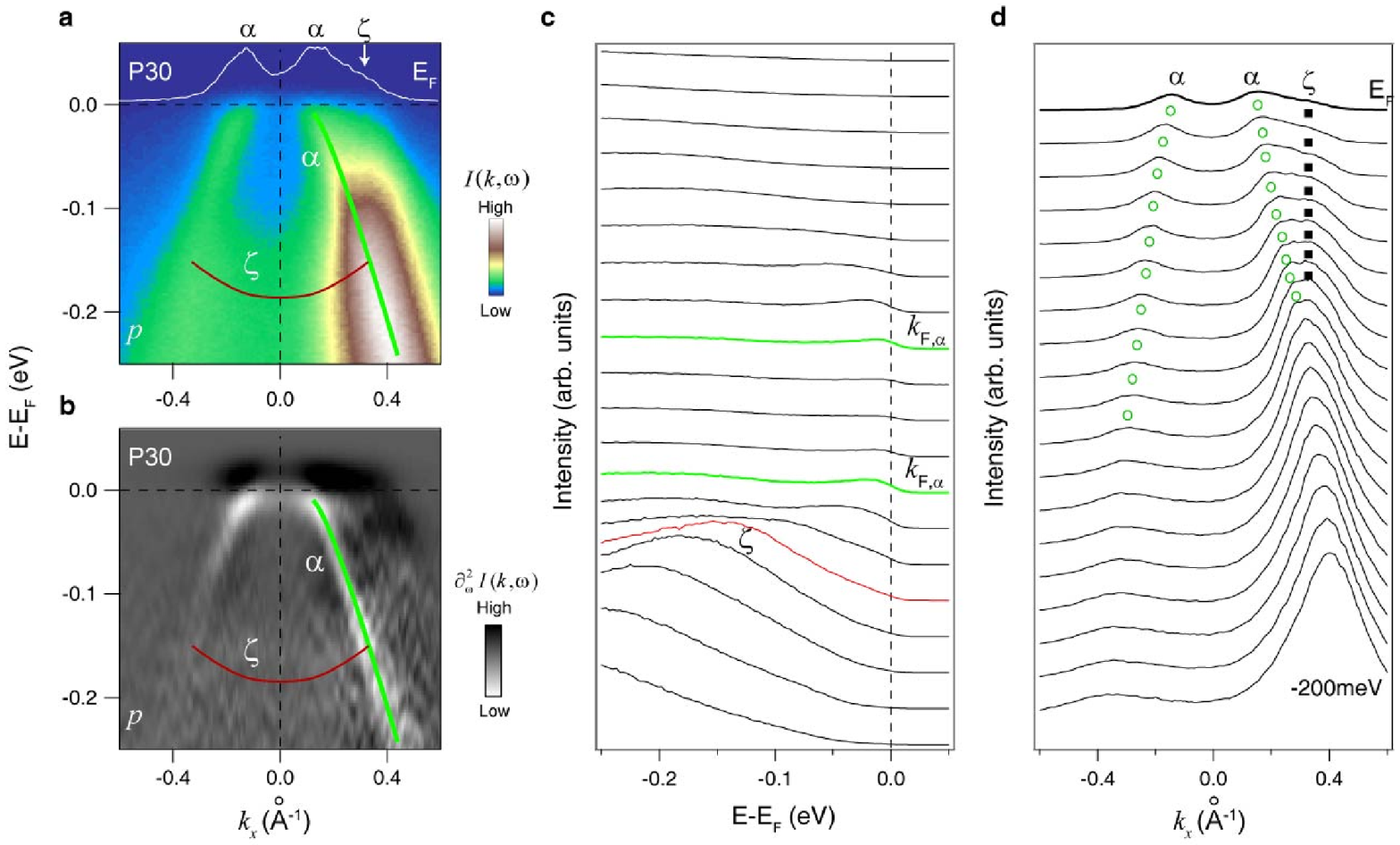}
\caption{\textbf{The photoemission data of P30 near $E_F$ taken in
the $p$ polarization.} \textbf{a}, \textbf{b}, The photoemission
intensity ($I(k,\omega)$) and its second derivative with respect to
energy ($\partial^2 I(k,\omega)/\partial\omega^2$) taken with 120~eV
photons around $\Gamma$ in the $p$ polarization, respectively. Solid
lines are band dispersion, as determined by momentum distribution
curve (MDC) peaks or the minima of the second derivative of
photoemission intensity with respect to energy. The inset in
\textbf{a} is the MDC at $E_F$. \textbf{c}, \textbf{d}, The energy
distribution curves (EDC's) and  MDC's of the data in \textbf{a},
respectively.} \label{dz2}
\end{figure*}

\newpage

\section{Photon energy dependence}

By changing the photon energy in ARPES experiment, one can probe the
different $k_z$ values, according to the following formula
\cite{hufner}:
\[\begin{array}{l}
 {k_\parallel } = \frac{{\sqrt {2m} }}{\hbar }{[(h\nu  - \Phi )]^{1/2}}\sin \theta  \\
 {k_z} = \frac{{\sqrt {2m} }}{\hbar }{[(h\nu  - \Phi ){\cos ^2}\theta  + {V_0}]^{1/2}} \\
 \end{array}\] where m and $\hbar$ are the electron mass and Planck¡¯s constant, $\nu$ is
the photon energy, $\Phi$ is the work function, and $V_0$ is the
inner potential of the sample, which could be determined
experimentally. Through matching the periodicity of the photon
energy dependence to the high symmetry points of the Brillouin zone
along $k_z$ direction, the inner potential ($V_0$) is set to $16~eV$
for calculating the $k_z$'s in our experiment.

Fig.~S5 shows the photon energy dependence of the MDC's near $E_F$.
The Fermi crossings of different bands could be traced separately in
the $s$ and $p$ polarization. The peak positions of the MDC's are
marked by combining the highest intensity and the center of mass,
which is a reliable way and is practiced commonly in ARPES
community. Note that, while the $\alpha$ band move outwards from
$\Gamma$ to Z, the even $d_{xz}$ orbital in the $\alpha$ band could
mix into the $\beta$ band. As shown in Figs.~S5e-h, when the $k_z$
moves away from $\Gamma$, two peaks could be identified
corresponding to the $\alpha$ and $\beta$ bands in the $p$
polarization. The intensity transfer between the MDC peaks of
$\alpha$ and $\beta$ observed in the $p$ polarization is due to the
$k_z$ dependence of the even orbital mixing between the $\alpha$ and
$\beta$ band.

\begin{figure*}[p!]
\includegraphics[width=13cm]{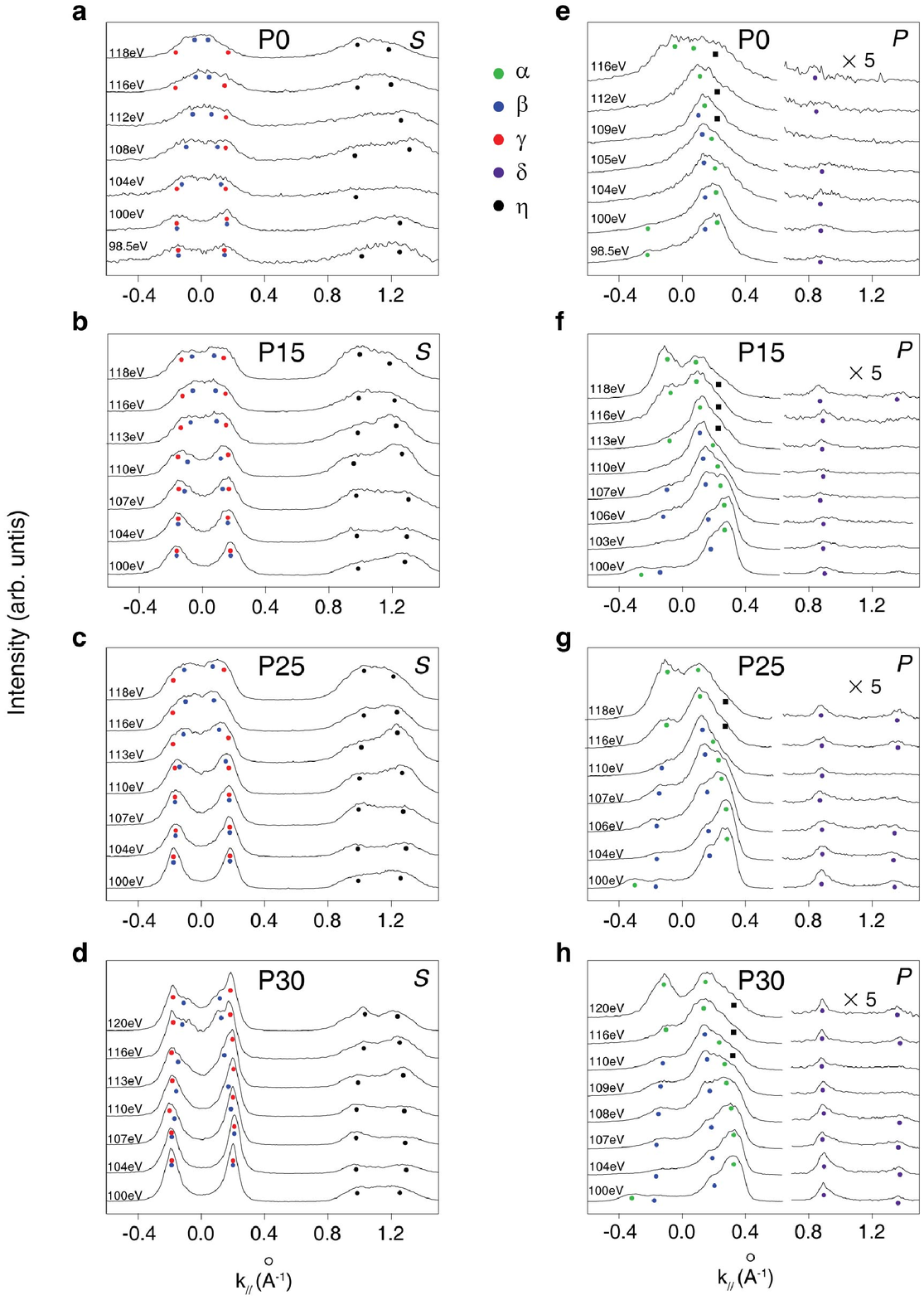}
\caption{\textbf{The photon energy and doping dependence of the
MDC's near E$_F$ in the $s$ and $p$ polarizations.} \textbf{a-d},
The photon energy dependence of the MDC's near E$_F$ taken in the
$s$ polarization in P0, P15, P25, and P30 samples, respectively.
\textbf{e-h}, are the same as \textbf{a-d}, but taken in the $p$
polarization. Note that the right parts of \textbf{e-h} are in an
expanded intensity scale. The peaks marked by black solid squares
are contributed by the residual spectral weight of the band below
$E_F$ with the $d_{z^2}$ orbital} \label{MDC}
\end{figure*}